\documentclass[journal]{IEEEtran}
\usepackage{cite}
\usepackage{amsmath}
\usepackage{graphicx}
\usepackage{url}
\usepackage{color}

\ifCLASSINFOpdf
\else
\fi
\begin{document}
	\title{Analysis of Avoided Transmission Through Decentralized Photovoltaic and Battery Storage Systems}
	
	%\author{Shigeyoshi~Sato,~\IEEEmembership{}
	%	Anke~Weidlich,~\IEEEmembership{}% <-this % stops a space
	%	\thanks{Anke Weidlich is with Albert-Ludwigs-Universit{\"a}t Freiburg.}% <-this % stops a space
	%	\thanks{Manuscript received August 1, 2017; revised Month XX, 201X.}}

	\author{\IEEEauthorblockN{Shigeyoshi Sato}
		\IEEEauthorblockA{Faculty of Environment and Natural Resources, \\
			University of Freiburg, Germany}
		\and
		\IEEEauthorblockN{\\Anke Weidlich}		
		\IEEEauthorblockA{Department of Sustainable Systems Engineering, \\
			University of Freiburg, Germany \\
			Email: anke.weidlich@inatech.uni-freiburg.de}
		\thanks{
			{\copyright} 2019 IEEE. Personal use of this material is permitted. Permission from IEEE must be obtained for all other uses, in any current or future media, including reprinting/republishing this material for advertising or promotional purposes, creating new collective works, for resale or redistribution to servers or lists, or reuse of any copyrighted component of this work in other works. 
			
			DOI 10.1109/TSTE.2019.2946446, 
			
			https://ieeexplore.ieee.org/document/8863430}
}
	

	\markboth{IEEE Transactions on Sustainable Energy (Accepted version)}%
	{Shell \MakeLowercase{\textit{et al.}}: Bare Demo of IEEEtran.cls for IEEE Journals}



	\maketitle
	
	\begin{abstract}
		Decentralized renewable energy systems can be low-carbon power sources, and promoters of local economies. It is often argued that decentralized generation also helps reducing transmission costs, as generation is closer to the load, thus utilizing the transmission system less. The research presented here addresses the question whether or not, or under what circumstances this effect of avoided transmission can actually be seen for a community-operated cluster of photovoltaic (PV) power plants in two sample locations, one in Germany and one in Japan. For the analysis, the newly developed instrument of MPI-MPE diagrams is used, which plot the maximum power import (MPI) and maximum power export (MPE) in relation to the reference case of no local generation. Results reveal that for moderately sized PV systems without battery storage, avoided transmission can be seen in the Japanese model location, but not in Germany. It was also found that an additional battery storage can lead to avoided transmission in both locations, even for large sizes of installed PV capacity.
	\end{abstract}
	

	% \fi
	\IEEEpeerreviewmaketitle

	\section*{Nomenclature}
	
	\noindent \begin{tabular}{ll}
		$\lambda_1,\lambda_2$ & Objective weighting factor \\
		$C$ & Nominal storage capacity of the battery (MWh) \\
		$Ch_t$	 & Charging power at time $t$ (MW) \\
		$DG_t$	& Discharging power to the grid at time $t$ (MW) \\
		$DS_t$ & Discharging power for self-consumption (MW) \\
		$t$ & Time interval (h or 15 min) \\
		$P^{\text{max}}$ &	Maximum power of grid interaction (MW) \\
		$RL_t$ & Residual load at time $t$ (MW) \\
		$SG_t$ & Surplus generation at time $t$ (MW) \\
		$S_t$ & Storage charge level at time $t$ (MWh) \\
	\end{tabular}

	\section{Introduction}
	
	Decentralized renewable power generation gains much attention as an
	environmentally friendly power source and a promoter of local economies.
	One additional advantage often advocated is that decentralized renewable
	electricity generation helps avoiding infrastructure cost, because the
	transmission system is used less, as generation is geographically close
	to the consumption \cite{Kahn2008}. However, in many cases, variable
	renewable energy hardly reduces the annual net peak load \cite{Ueckerdt2013}. The authors of \cite{VonAppen2015} quantitatively illustrate that grid interaction can decrease with the introduction of on-site photovoltaic (PV) systems in Australia, but show the effect only for one case with a fixed PV size. With larger PV sizes, increased generation could cancel
	out the avoided transmission and even require an enhancement of grid
	infrastructure for exporting the surplus electricity. Therefore, it is
	interesting to study how avoided transmission depends on the PV size,
	and how it changes if battery storage is added to the local energy
	system.
	
	The potential of combinations of PV systems with battery storage has
	been widely studied. \cite{Luthander2015,Merei2016,Braam}, among others, show how cost savings can be achieved, and self-consumption rates are increased with larger battery systems operated along with PV plants. The benefit of PV and battery systems for	distribution system operators (DSOs) due to peak shaving has been
	demonstrated by \cite{Li2014,BolivarJaramillo2016,buedenbender}, among
	others. Several studies also analyzed the power flow at the distribution
	transformer for evaluating the grid interaction induced by distributed
	PV system, e.\,g. \cite{buedenbender,Nykamp2012,Zupancic2020}. The authors of \cite{VonAppen2015} and \cite{Riesen2017} focused their	studies on single households, \cite{Venu2009} and \cite{Klaassen2015} looked at residential areas, and \cite{Nykamp2012} studied an entire village of prosumers with PV systems. \cite{Klaassen2015} specifically address the relationship between maximum power and PV size, which is not reported in most other studies. They derive values of annual maximum residual load (MRL) and annual maximum surplus generation (MSG) from net load duration curves of three residential areas equipped with different sizes PV generators. Similar approaches had been previously followed by \cite{Ueckerdt2013} and \cite{Schill2014}, but for studying the effects of PV usage at a
	country-wide level.
	
	This work aims at comprehensively estimating the impact of
	decentralized PV systems on transmission grid usage. A PV plant was
	chosen as a representative system for renewable power generators, because PV
	has been intensively studied \cite{Ogbomo2017,Sato}, and
	is widely implemented world-wide. Two model communities have been analyzed which have very different load and solar irradiation patterns. Two cases were studied for comparison, i.\,e. PV-only systems and PV systems combined with a battery storage. The marginal change in transmission flows related to the studied systems is not directly evaluated. Instead, the power exchange at the transformer connecting the model communities to the public grid is taken as a proxy for transmission requirements related to serving the community's electricity demand. 
	
	The work builds on the concepts of MRL and MSG introduced by \cite{Klaassen2015}, and uses them to form the newly proposed MPI-MPE diagrams, which plot the annual maximum power import (MPI, in MW) and annual maximum power export (MPE, in MW) that result from the introduction of a local PV system with or without an additional battery storage into the community's local grid. This new instrument is a very	insightful and yet simple visualization of the effect that different PV sizes have on the exchange of power at the community's connection point to the public grid. Through this, transmission reductions from
	decentralized generation can be analyzed over a range of PV sizes in an
	intuitive manner. To the best knowledge of the authors, similar tools do
	not exist in the literature, so the proposed MPI-MPE curves extend the
	state-of-the-art of transmission reduction analysis for the domain
	studied here.
	
	\section{The Method}
	
	The power exchange of a community at a distribution transformer that connects it to the public power grid is analyzed here. The higher of the two values MPI and MPE as introduced before is taken as an indicator of how much the community uses the upstream (including the transmission) system for satisfying its electricity demand. If local generation reduces the maximum power exchange, this is referred to as avoided transmission. It is argued that the cost of transmission is mainly influenced by the infrastructure investment, and less by its operation. Therefore, regarding the maximum power exchange rather than the total energy exchanged is most relevant, as infrastructure is usually dimensioned so that it can always serve the peak load, or absorb the peak surplus generation.
	
	The load consists of private households, industrial consumers, a few agricultural farms
	and further consumers that form a model community. The community can always draw any required power from the public grid and also feed its surplus power into the grid. In Case 1, on-site PV modules are assumed to be installed. In Case~2, PV plants are installed along with a centralized battery (or several decentralized batteries that are centrally controlled and can therefore also be modeled as one unit). In the latter case, the storage capacity is varied in the range of 1.5\,--\,4.5\,kWh storage capacity per kW$_\text{p}$ of installed PV power (here referred to as kWh/kW$_\text{PV}$). 
	
\begin{figure}[h]
		\centering
		\includegraphics[width=0.45\textwidth]{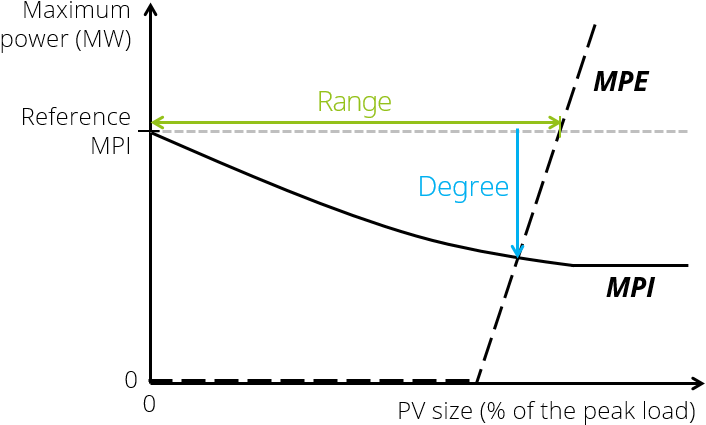}
		\caption{MPI-MPE diagram}
		\label{MPEMPI}
	\end{figure}

	MPI and MPE are calculated on the basis of the net load, i.\,e. the load minus the PV generation. The net load is first separated into positive values -- the residual load RL -- and negative values -- the surplus generation SG -- of the community. Of this, following \cite{Klaassen2015}, the annual maximum RL value constitutes the MRL, and the annual maximum SG value is the MSG (both in MW). These values are then calculated for various sizes of the PV modules. With the values computed, the proposed MPI-MPE diagram can be developed. Such a diagram is exemplarily depicted in Fig.~\ref{MPEMPI}. In this graph, the solid line indicates MPI and the dotted line indicates MPE at a certain PV size. On the abscissa, the PV size is indicated in percent of the peak load. The range indicates the PV size for which transmission reductions are achieved. It is limited by the intersection of MPE and the power exchange with the feeder system when no local generation is present. The latter is marked as reference MPI in Fig.~\ref{MPEMPI}. The degree is the vertical distance between the reference MPI and the intersection of MPE and MPI. It indicates the decrease in maximum power exchanged at the transformer which can be achieved at best for an ideal PV size.
	
	In Case~1, MRL and MSG are equal to the annual MPI from and MPE to the public grid, respectively. In Case~2, optimum battery operation is determined through a multi-objective linear program (MOLP), following a similar approach by \cite{Riesen2017}. The MOLP is solved for the scheduling horizon $T$ of one year applying a rolling horizon approach \cite{Silvente2015} with a	control horizon of 24 hours and a prediction horizon 144 hours. The 24-hour control horizon always begins at 9:00\,a.m. of a day. It was found that the optimum computed for an overall horizon of one week (168 hours) does not differ from that for longer periods, so this constitutes the
	chosen number of time steps analyzed in each step of the rolling horizon
	procedure. 
	
	The input data are the load and PV generation profiles for
	one year in the target community. They form the two time series of
	residual load (positive net load), $RL_t\geq 0~\forall t = 1,2,...,T$, and surplus generation (absolute values of negative net load), $SG_t\geq 0~\forall t=1,2,...,T$, for time steps $t$. These time series are based on historical data, so perfect foresight is assumed here, and forecast uncertainty is neglected. 
	
	The objective function is defined as the minimization of the maximum absolute power exchanged between the community and the public grid, $P^{\text{max}}$ (Eq.~\ref{objective}). In the objective function, $\lambda_1$ and $\lambda_2$ are objective weighting factors, and $0\leq\lambda_1,\lambda_2\leq 1$. The objective function defined here discourages simultaneous battery discharging to the grid and charging. The given model with $\lambda_1=10^{-3}$ and  $\lambda_2= 10^{-6}$ puts high emphasis on the main goal to minimize the absolute power exchange, while completely avoiding simultaneous charging and discharging.
	
	The battery is characterized by its nominal capacity $C$ (in MWh) and its state-of-charge $S_t$ (in MWh), with the initial storage level $S_0$. The state-of-charge can take any value between 0.1 and 0.9 times the storage capacity (\ref{battcharrange}), which avoids high battery aging \cite{Li2014,Merei2016}.%,Bost2014}. 
	
	The charging/discharging power at time $t$, $Ch_t$, can take any value of surplus generation (\ref{battchar}). No additional technical constraints on charging and discharging power are assumed. Although this may seem to overestimate avoided transmission, optimization outputs show that the highest charging rates were observed for the largest batteries reported in Sections~\ref{sec:CaseStudy} and \ref{sec:results}, and that the observed power-to-energy ratios of the battery in these cases were in the range of 0.15\,--\,0.17\,kW/kWh. Many battery systems currently available on the market offer some flexibility in choosing the maximum charging/discharging power in relation to the energy storage capacity. An extensive market overview of industry-scale batteries in \cite{PVmag} shows that the majority of standard configurations for  offered battery systems are in the range of 0.2\,--\,1.5\,kW/kWh, with less than 10\,\% having a power-to-energy ratio below 0.2. Therefore, we assume that charging/discharging power is not a limiting factor for the battery sizes considered in the current work. Similar to \cite{Zeh2014} and \cite{Schneider2014}, no constraints were introduced that limit battery discharge to the public grid. Consequently, the battery can be discharged even beyond local demand, typically at night, so as to free storage capacity for the next surplus period, which usually happens during summer days \cite{Braam}. However, in order to minimize grid interaction, battery discharging for feeding it into the grid at time step $t$, $DG_t$, is in the following distinguished from discharging for self-consumption, $DS_t$. This allows for minimizing discharge to the grid while encouraging the usage of stored energy for self-consumption. 
	
	Energy losses during charging and discharging are neglected. The resulting energy balance is given by (\ref{battSOC}), where $\Delta t$ is the duration of one time interval. 
	
	(\ref{maxP1}) and (\ref{maxP2}) limit both imports from the grid, $RL_t-DS_t$, and exports to the grid, $SG_t-Ch_t+DG_t$,  to $P^{\text{max}}$, which is minimized in the objective function. The resulting optimization problem is formulated as follows: \begin{equation}\label{objective}
	\min z = \left(1-\lambda_1-\lambda_2\right) P^{\text{max}} -\lambda_1 \sum_t DS_t+\lambda_2\sum_t DG_t
	\end{equation} s.\,t. (all constraints $\forall t=1,2,...,T$) \begin{equation}\label{battcharrange}
	0.1\leq S_t\leq 0.9
	\end{equation} \begin{equation}\label{battchar}
	Ch_t \leq SG_t
	\end{equation} \begin{equation}\label{battSOC}
	S_t=S_{t-1}+\left(Ch_t-DS_t-DG_t\right)\cdot \Delta t
	\end{equation} \begin{equation}\label{maxP1}
	RL_t-DS_t\leq P^{\text{max}}
	\end{equation} \begin{equation}\label{maxP2}
	SG_t-Ch_t+DG_t\leq P^{\text{max}}
	\end{equation} \begin{equation}\label{nonnegative}
	P^{\text{max}}, DG_t, DS_t, Ch_t\geq 0
	\end{equation}
	
	The problem was solved with the Matlab function linprog and takes in the order of one minute to compute per PV/battery size combination. Finally, MPI and MPE are calculated from the optimization outcome as the maximum residual load after storage discharging, and the maximum net surplus after charging, respectively (\ref{MPI}, \ref{MPE}). \begin{equation}\label{MPI}
	MPI=\max(RL_t-DS_t)
	\end{equation} \begin{equation}\label{MPE}
	MPE=\max(SG_t-Ch_t + DG_t)
	\end{equation}

	\section{Case Study}\label{sec:CaseStudy}
	
	In this study, two model communities were considered. The first one is
	situated in the medium-sized city Erding in Germany. The second one is a
	similarly structured community in the region of Shikoku in Japan. Germany and
	Japan were chosen as showcases, because they have the first and the
	second largest PV penetration per capita in the world \cite{IEA}.
	Erding and Shikoku were chosen due to their comparable population
	density, and for their comparable economic structure, with an almost same
	distribution of gross value added among economic sectors \cite{EU,Shokosoken}. The model communities can be categorized as suburb areas where farms, factories, shops, offices, and
	residential buildings are mixed.
	
\begin{figure}[h]
		\centering
		\includegraphics[width=0.5\textwidth]{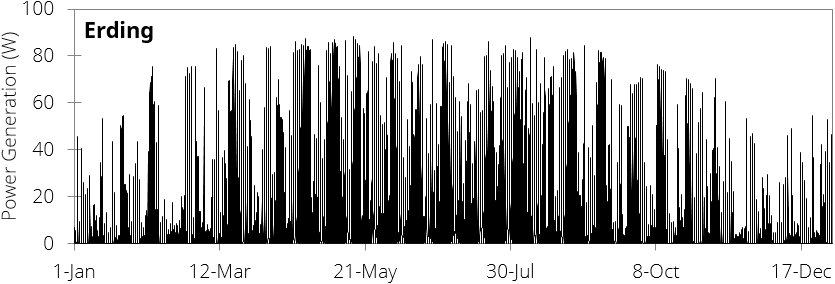}
		\includegraphics[width=0.5\textwidth]{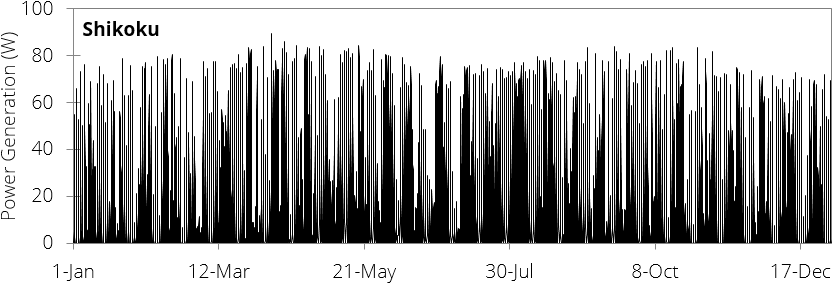}
		\caption{Power generation profiles in Erding (top) and Shikoku (bottom) for a 100\,W PV plant}
		\label{fig:genprofile}
	\end{figure}

	Load data was obtained from the local DSO companies \cite{SWErding,SWShikoku}. The time resolution is 15~min for Erding, and one hour for Shikoku. The peak load recorded in Erding is 36.22\,MW. Data of Shikoku was scaled to have the same peak load as Erding. Following the approach of \cite{Lovelace2015} and \cite{Regalado2016}, the PV generation profile for both locations were simulated with TRNSYS, using type194, which is a one-diode, five parameter model as developed by \cite{DeSoto2006}, and applying tilted angles of the PV panels of 35$^\circ$ for Germany and 30$^\circ$ for Japan. For weather data, the reference year based on measurements between 1995 and 2012 for Erding was taken from \cite{DWD}. For Shikoku, data was obtained from \cite{NEDO} (for the location Takamatsu), which is based on measurements between 1990 and 2009. The resulting 15\,min / hourly power generation profiles of single PV modules (100\,W) in the year 2017 are shown in Fig.~\ref{fig:genprofile}. It has to be mentioned that the time resolution determines the degree of precision for the absolute peak of power exchange from/to the community. Higher absolute power values can occur within one 15\,min or 1\,h period, while the given data only provides average power per time interval. However, as a whole community is considered here, load curves are smoother than for single consumers' curves, therefore the time resolution is considered sufficient for the analysis provided here.
	
	The generation profiles reveal noticeable differences between the two
	considered locations. The profile in Erding has greater seasonal
	variation, with less power output occurring during winter. The generation in
	Japan, in contrast, is more evenly distributed throughout the year. During summer, daily PV generation peaks are often higher in Erding than in Shikoku, which is due to the different temperature levels in both locations that have an effect on PV performance. In addition, thanks to longer lengths of daytime in German summer, PV plants can harvest more sunlight per day, so more electricity
	is generated during an average summer day in Erding than in Shikoku.
	
	\begin{table*}[t]
		\centering
		\begin{tabular}{|r|r|r|r|r|r|}
			\hline
			\multicolumn{2}{|c|}{\textbf{Community load}} & \multicolumn{2}{|c|}{\textbf{100\,\% capacity size}} & \multicolumn{2}{|c|}{\textbf{100\,\% energy size}} \\
			\hline
			\textbf{Peak power} & \textbf{Energy demand} & \textbf{Peak power} & \textbf{Energy generation} & \textbf{Peak power} & \textbf{Energy generation} \\
			\hline
			\multicolumn{6}{|l|}{\textbf{\emph{Erding}}} \\	
			\hline			
			36.22\,MW & 183.8\,GWh & 36.22\,MW$_\text{p}$ & 38.37\,GWh & 173.6\,MW$_\text{p}$ & 183.8\,GWh \\
			& & (31.98\,MW actual peak) & & (= 480\,\% of the peak load; & \\
			& & & & 153.3\,MW actual peak) & \\
			\hline
			\multicolumn{6}{|l|}{\textbf{\emph{Shikoku}}} \\
			\hline
			36.22\,MW & 196.8\,GWh & 36.22\,MW$_\text{p}$ & 46.05\,GWh & 154.8\,MW$_\text{p}$ &  196.8\,GWh\\
			& & (nominal capacity; & & (= 428\,\% of the peak load;  &  \\
			& & 32.33\,MW actual peak) & & 138.2 MW actual peak) & \\
			\hline
		\end{tabular}
		\caption{Power and energy balances in Erding and Shikoku}
		\label{tab:load}
	\end{table*}
	
	The integral over the power profile shows a generation of 106\,kWh in Erding and 127\,kWh in Shikoku, respectively, from a PV system with nominal capacity of 100\,W$_\text{p}$. Power and energy values for other plant sizes are summarized in Tab.~\ref{tab:load} for both locations. In this table, the nominal capacity of the photovoltaic modules (in W$_\text{p}$) is expressed in two different ways. One definition puts the plant size in terms of percent of the peak load, following \cite{Merei2016}. Consequently, the PV size of 100\,\% is equal to the community's peak load of 36.22\,MW for both communities. Although the peak PV generation does not necessarily occur at the same time as peak consumption, this is a helpful value which can easily be determined for any community considered. The second definition takes the yearly aggregate PV generation as the reference. Following this reasoning, the 100\,\% energy size equals the PV plant size that allows a yearly aggregate generation equal to the community's yearly aggregate electricity consumption. It is observed that the latter PV size is 480\,\% of the peak load in Erding, and 428\,\% in Shikoku. This measure is, again, a helpful reference size which can easily be computed for any location. In all discussions that follow below, PV size values are always expressed in reference to the capacity size (i.\,e. peak load), if not stated otherwise.
	
	\section{Results and Discussion}\label{sec:results}
	
	In the following, results are presented for the two considered cases
	individually. Findings for each case are compared between the two model
	locations Erding and Shikoku.
	
	\subsection{Case 1 (PV Only)}\label{results-case-1-pv-only}
	
\begin{figure}[h]
		\centering
		\includegraphics[width=0.5\textwidth]{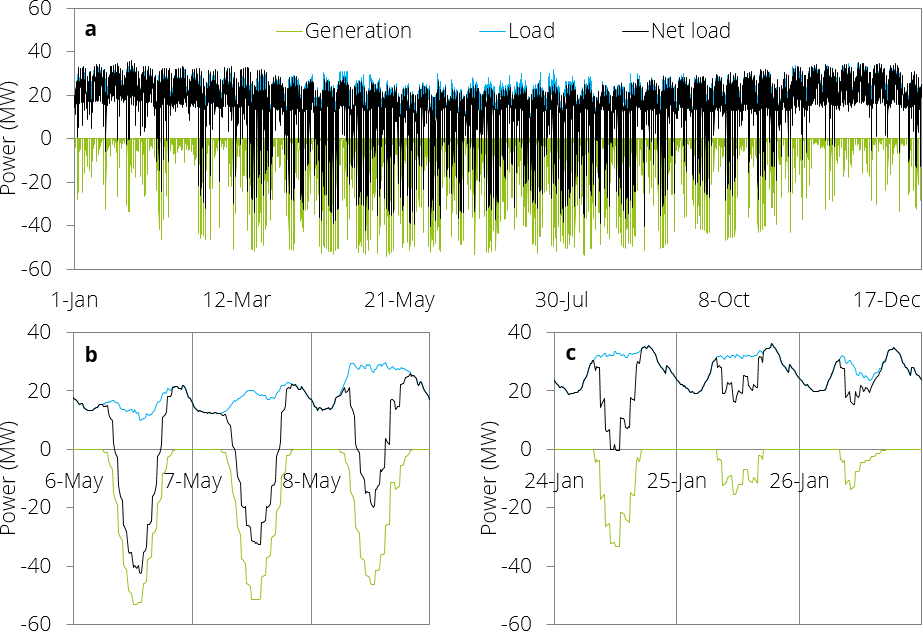}
		\caption{Power profiles for Case~1 (PV size 169\,\%) in Erding for a) whole year, b) May, c) January}
		\label{fig:powerprofileErding}
	\end{figure}

	For the analysis of Case~1, the net power profile was calculated by
	subtracting PV generation from electrical load. The net load in Erding is depicted in Fig.~\ref{fig:powerprofileErding} for an example PV size. Peak load occurs during winter, and the annual maximum of 36.22\,MW was observed at 6:30\,p.m. on 25$^{\text{th}}$\,January. No PV generation occurs at the time of the peak load (cp. Fig.~\ref{fig:powerprofileErding}c). Fig.~\ref{fig:powerprofileErding}b shows MSG values, with a maximum on 6$^{\text{th}}$\,May. MSG (42.4\,MW) is higher than MRL, therefore grid interaction is increased in comparison to the reference situation without PV.
	
\begin{figure}[h]
		\centering
		\includegraphics[width=0.5\textwidth]{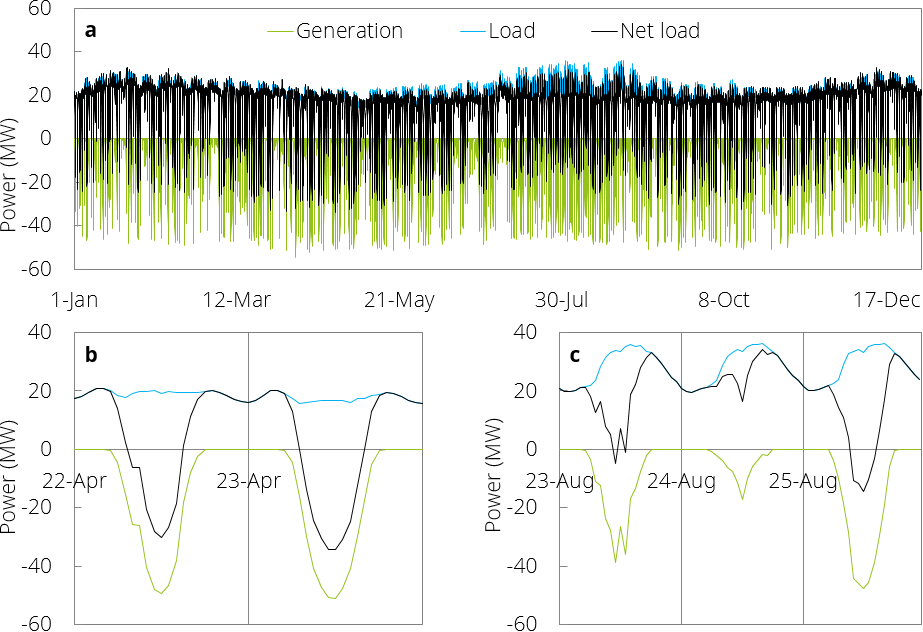}
		\caption{Power profiles for Case~1 (PV size 169\,\%) in Shikoku for a) whole year, b)  April, c) August}
		\label{fig:powerprofileShikoku}
	\end{figure}

	In Shikoku, grid interaction is lowest for a PV size of 169\,\%, as it will be shown later. Fig.~\ref{fig:powerprofileShikoku} depicts the net load in Shikoku for this plant size. The peak load of 36.22\,MW was observed at 4:00\,p.m. on
	24$^{\text{th}}$\,August. Fig.~\ref{fig:powerprofileShikoku}c shows that during peak load, the PV plants deliver part of the supply, leading to a reduced MRL value of 34.33\,MW. On the other hand, surplus was also observed, indicating exports to the public grid. The annual absolute maximum value of SG, as depicted in Fig.~\ref{fig:powerprofileShikoku}b, appeared at 11:00 on 23$^{\text{rd}}$\,April, with an MSG of 34.16\,MW. Overall, grid interaction was reduced by 5.2\,\% (from 36.22\,MW to 34.33\,MW).
	
	In order to analyze the dependency of MSG and MRL on the PV size, ordered duration curves of the net load for different PV sizes are plotted in Fig. 5. The leftmost values of the curves indicate the MRL, while	the rightmost values give the MSG for each analyzed size. As Case~1 involves no further power control, MRL and MSG correspond to MPI and MPE values, respectively. It can be observed that as PV size increases, MPI decreases only slightly in Shikoku, but MPE rises quickly. In Erding, MPI does not change at all with increasing PV sizes, and MPE
	decreases in a similar way as in Shikoku.

	\begin{figure*}[t]
		\centering
		\includegraphics[width=0.84\textwidth]{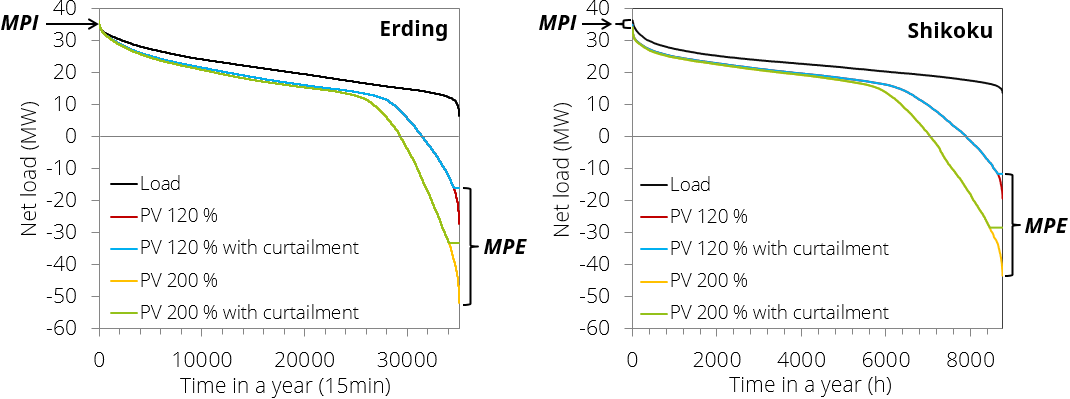}
		\caption{Duration curves of net load for Case~1 in Erding (left) and Shikoku (right)}
		\label{fig:loadduration}
	\end{figure*}

	Fig.~\ref{fig:loadduration} also depicts the case of curtailment, which is discussed as an appropriate method for congestion management \cite{Joos2018,KlingeJacobsen2012}. The
	easiest implementation of curtailment is a fixed limit given as the
	percentage of the installed capacity up to which power can be fed into
	the grid. Any feed-in above the limit is curtailed. This is referred to
	as ``static curtailment'' \cite{Wieben2016a,Bongers2016,Wiest2017} or ``fixed curtailment'' \cite{Rossi2016}. An alternative implementation is ``dynamic curtailment / approach''. In the dynamic approach, generators are only curtailed in situations in which they actually contribute to grid congestion. While the static approach requires no information about the current network state, the dynamic approach requires a communication
	system, because PV generators have to receive the information about
	current grid state continuously. For simplicity, the static approach is
	considered here. It specifies that up to 5\,\% of annual energy which comes at the highest power can be curtailed. If this is applied, then MPE significantly decreases in comparison to the initial MSG, while MPI is equal to MRL.
	
\begin{figure}[h]
		\centering
		\includegraphics[width=0.34\textwidth]{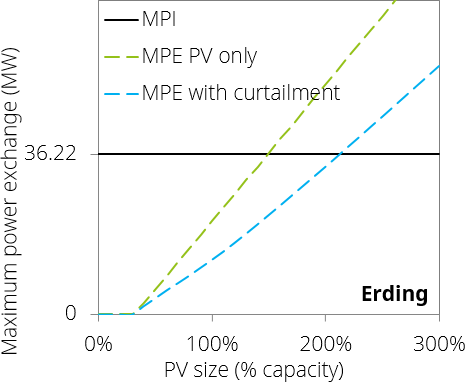}
		\includegraphics[width=0.34\textwidth]{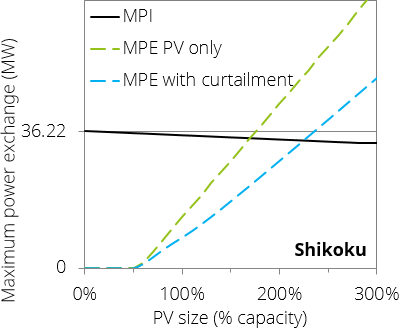}
		\caption{MPI-MPE diagram of Case~1 in Erding (top) and Shikoku (bottom)}
		\label{fig:MPIMPE}
	\end{figure}

	The relationship between maximum power peaks and PV size was extracted
	from the duration curves and summarized in Fig.~\ref{fig:MPIMPE} to form the MPI-MPE
	diagram. In	Erding, the maximum power flux to/from is the same for any PV size
	between 0 and 149\,\%, without curtailing assumed. For any larger PV
	size, MPE increases above the reference of no PV generation. For Shikoku, it shows that the maximum power flux to/from the grid is less than 36.22\,MW for a range of PV sizes up to 175\,\% without curtailment, and up to 235\,\% with the described 5\,\% curtailment. With no curtailment, the grid interaction reaches a minimum at a PV size of
	169\,\%, corresponding to a 5.2\,\% reduction in relation to the reference setting
	without PV. In the case of curtailment, the grid interaction
	is minimized at a PV size of 224\,\%, where both MPI and MPE are 33.72\,MW. This corresponds to a 6.9\,\% reduction of grid interaction. Thus, curtailment reduces transmission in terms of both range and degree. 
	
	In summary, the effect of avoided transmission can be observed in
	Shikoku, but not in Erding, when only PV generators are installed. The difference between the two locations can be attributed to the degree of coincidence of the load
	peak and PV generation. The daily load peaks in Erding tend to appear at the later afternoon between 5:00 and 7:00\,p.m. in the winter season. At these times, the PV plants do not generate any power. In Shikoku, in contrast, the daily load peaks tend to appear in the early afternoon between 2:00 and 4:00\,p.m. in the summer season. This coincides well with the PV generation profiles (Fig.~\ref{fig:powerprofileErding}c).

	\subsection{Case 2 (PV and Battery)}\label{results-case-2-pv-and-battery}
	
	When battery storage is added to the PV systems, grid interaction decreases. The energy storage capacity of the battery is here quantified in relation to the PV peak capacity in kWh/kW$_\text{PV}$, as introduced earlier. Different PV sizes were combined with battery sizes of 1.5, 2.5, 3.5 and 4.5\,kWh/kW$_\text{PV}$. It was observed that the configurations that lead to the lowest power exchange with the public grid were at PV sizes of 426\,\% in Erding and 387\,\% in Shikoku, for the largest considered battery size of 4.5\,kWh/kW$_\text{PV}$. In the following, theses specifications are described in more detail for an illustration of the results.
	
\begin{figure}[h]
		\includegraphics[width=0.5\textwidth]{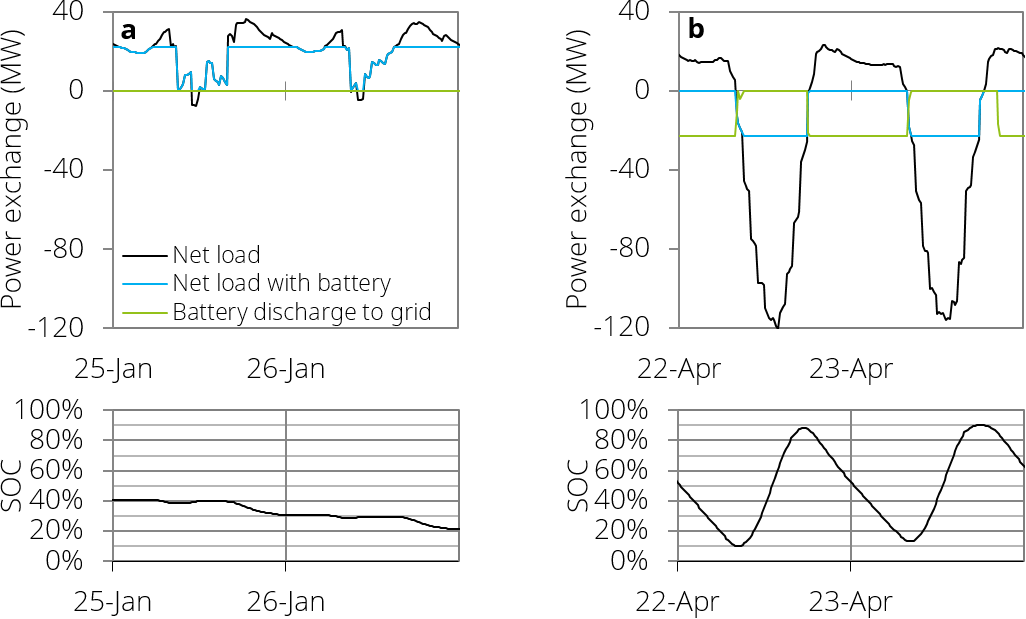}
		\includegraphics[width=0.5\textwidth]{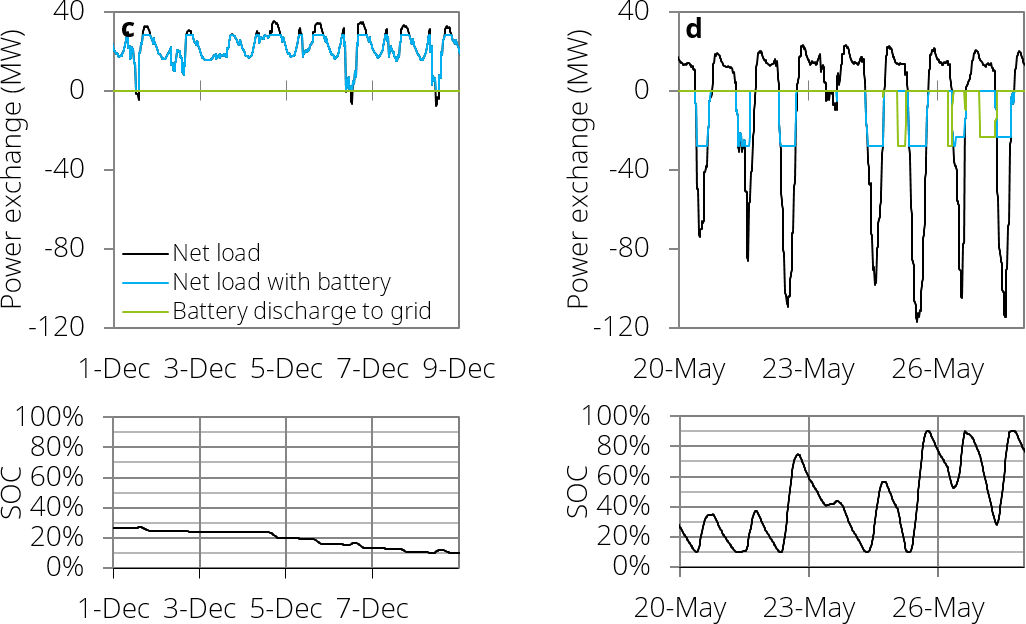}
		\caption{Power profiles for Case~2 (PV size 426\,\%, battery size 4.5\,kWh/kW$_{\text{PV}}$) in Erding for the days when a) MRL, b) MSG, c) MPI with battery, and d) MPE with battery appeared}
		\label{fig:powerProfileBatteryErding}
	\end{figure}

	In Erding, a 4.5\,kWh/kW$_{\text{PV}}$ battery at a PV size of 426\,\% has a storage capacity of 694\,MWh (cp. numbers in Tab.~\ref{tab:load}; $36.22\cdot426\%\cdot4.5=694$). The annual net load profiles with and without battery are summarized in Fig.~\ref{fig:powerProfileBatteryErding}, along with the profile of discharging power to the grid ($DG_t$) and the course of the state-of-charge $SOC_t$. During the period shown in Fig.~\ref{fig:powerProfileBatteryErding}d, the whole allowable range from 10\,\% to 90\,\% of SOC was utilized. In contrast, the battery was almost empty during the period shown in Fig.~\ref{fig:powerProfileBatteryErding}c, which can be attributed to the scarcity of PV generation in winter.
	
\begin{figure}[h]
		\includegraphics[width=0.5\textwidth]{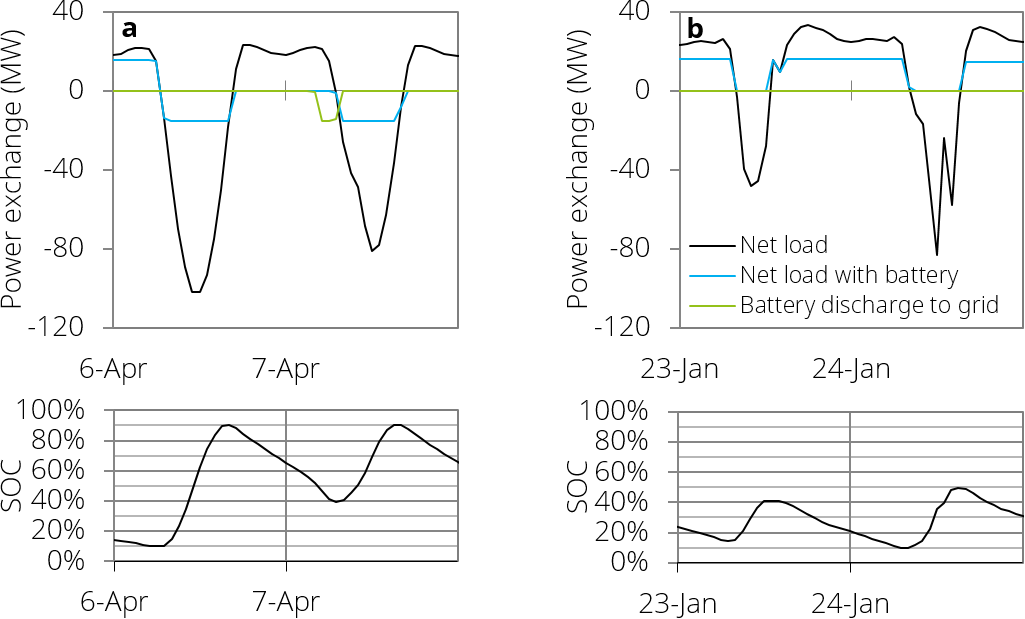}
		\includegraphics[width=0.5\textwidth]{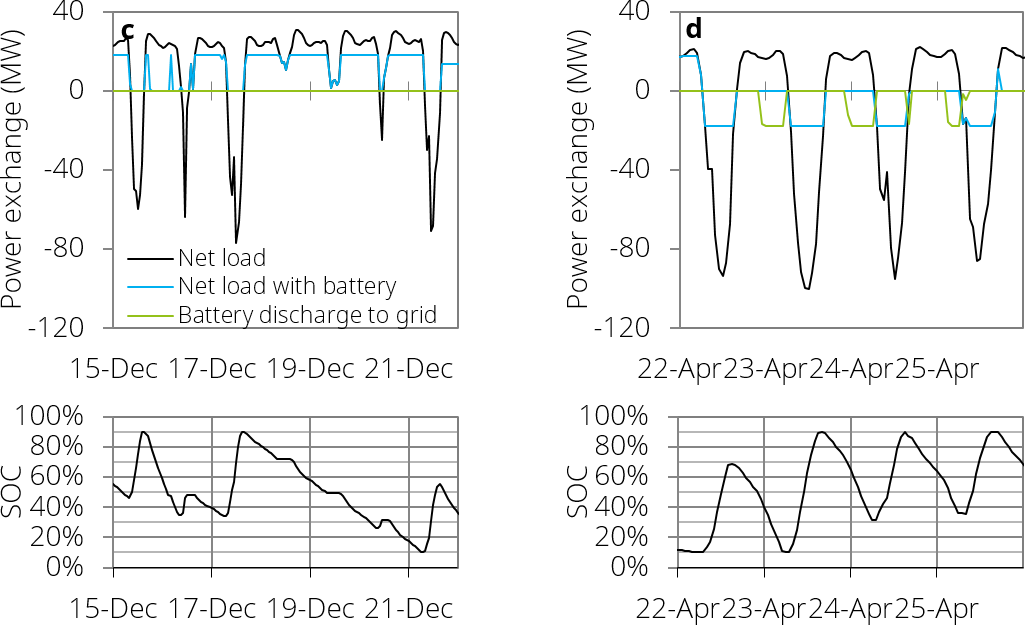}
		\caption{Power profiles for Case~2 (PV size 387\,\% battery size 4.5\,kWh/kW$_{\text{PV}}$) in Shikoku for the days when a) MRL, b) MSG, c) MPI with battery, and d) MPE with battery occurred}
		\label{fig:powerProfileBatteryShikoku}
	\end{figure}

	In Shikoku, a 4.5\,kWh/kW$_{\text{PV}}$ battery at a PV size of 387\,\% has a storage capacity of 631\,MWh. Power and SOC profiles for this configuration are summarized in Fig.~\ref{fig:powerProfileBatteryShikoku}. It was observed that the peaks in residual and surplus power profile are considerably reduced in the case with battery included, as compared to the PV-only cases. The timing of the highest power values also changed: While MRL and MSG for PV alone occurred in the end of January and beginning of April, respectively, MPI and MPE in the case with battery now appeared in the middle of December and end of April, respectively. MRL (33.01\,MW) at 6:00\,p.m. on 23$^{\text{rd}}$\,January was reduced through battery discharging (cp. Fig.~\ref{fig:powerProfileBatteryShikoku}a), and MSG (102\,MW) at 12:00\,a.m. on 6$^{\text{th}}$\,April was reduced by battery charging (cp. Fig.~\ref{fig:powerProfileBatteryShikoku}b). However, not all peaks were eliminated. Some time intervals of high positive or negative net load remain even with a battery of 631\,MWh. The value of MPI with battery is 17.96\,MW, which occurs during nighttime in mid-December, as shown in Fig.~\ref{fig:powerProfileBatteryShikoku}c. In contrast, Fig.~\ref{fig:powerProfileBatteryShikoku}d illustrates that MPE with battery was 17.92\,MW, which was observed during daytime towards the end of April. Overall, the introduction of a 631\,MWh battery decreased the grid interaction considerably from 102\,MW to 17.92\,MW. 
	
\begin{figure*}[ht]
		\centering
		\includegraphics[width=\textwidth]{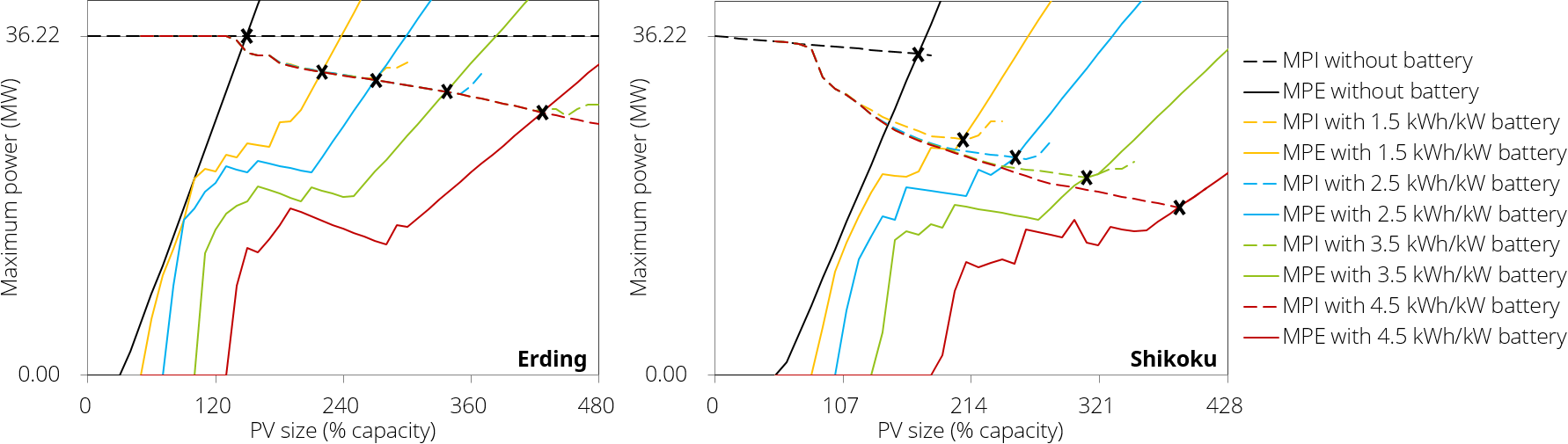}
		\caption{MPI-MPE diagram for Case~2 in Erding and Shikoku with various battery-to-PV size ratios}
		\label{fig:variedMPIMPE}
	\end{figure*}

	The dependency of MPI and MPE on PV and battery size is now further investigated. Fig.~\ref{fig:variedMPIMPE} shows the MPI-MPE diagram for five different storage sizes. The black crosses indicate the point of minimum grid interaction for each battery size. The right ends of the abscissas (PV size) are 480\,\% for Erding and 428\,\% for Shikoku, corresponding to the respective 100\,\% energy sizes (cp. Tab.~\ref{tab:load} and its explanation). It can be observed that MPI curves decrease and MPE curves	increase with larger PV size, until their point of intersection. Also, both MPI and MPE	curves decrease with larger storage capacity, indicating growing transmission avoidance. While MPI decreases strictly monotonously before, and MPE increases strictly monotonously after the intersection point with growing PV size, the two curves do not show a monotonous characteristic at the respective other side of the intersection. This is because the higher absolute value of the two, MPI and MPE, is the constraining factor. It is therefore not important in terms of objective function to also minimize the lower of the two values, so battery operation may be quite different for one PV size compared to another.
	
	For the battery sizes investigated, the minimum grid interaction was 17.92\,MW in Shikoku -- which is less than half the value as in the case without PV -- and 28.08\,MW in Erding.	In summary, it can be stated that both the degree and range of transmission avoidance proved to be smaller in Erding than in Shikoku for all battery sizes. Fig.~\ref{fig:variedMPIMPE} shows, e.\,g., that a 3.5\,kWh/kW$_\text{PV}$ battery yields an avoided transmission range of 380\,\% PV capacity in Erding, while that range is beyond the 100\,\% energy size in Shikoku (i.\,e. 428\,\% capacity). This indicates that in Shikoku, smaller batteries allow integrating more PV without transmission enhancement needs compared to Erding. The difference between the two locations lies in the seasonal distribution of PV generation profiles. In Shikoku, PV generation is more evenly distributed through the year (cp. Fig.~\ref{fig:genprofile}), so sufficient generation is available even in winter, and it can be used	to shave the RL peaks in winter (Fig.~\ref{fig:powerProfileBatteryShikoku}c). In contrast, PV generation in summer is lower than in Germany (Fig.~\ref{fig:genprofile}) due to shorter daytimes and higher temperature. Therefore, SG peaks are less pronounced in Shikoku (Fig.~\ref{fig:powerProfileBatteryShikoku}d), resulting in a smaller MPE value. On the other hand, Erding has larger seasonal variation and consequently faces a lack of PV generation in winter, as it can be observed in Fig.~\ref{fig:powerProfileBatteryErding}c. Also, because PV generation in summer is intensive (Fig.~\ref{fig:genprofile}), $SG$ peaks cannot easily be shaved (Fig.~\ref{fig:powerProfileBatteryErding}d).
	
	Relating the findings from Case~2 to the discussion of curtailing in Case~1 (cp. Fig.~\ref{fig:loadduration}), it must be stated that both options -- battery storage and curtailment -- avoid transmission in terms of range and degree. While curtailing comes at the cost of loosing a small part of the PV generation, batteries require additional investment. Costs must be set in relation to the savings that could be achieved through avoided transmission. The latter is very difficult to estimate; if only the transformer cost is considered, halving the initial size in the Shikoku example could reduce investment cost by an order of 200\,kEUR \cite{ACER} if the transformer is newly built or replaced. The investment into a battery of around 630\,MWh storage capacity from the same Shikoku example involves investment costs of an order of 100\,MEUR \cite{Bloomberg}. Even if additional benefits through increased self-consumption are included, the battery solution is an effective, but enormously expensive option for avoiding transmission.
	
	\section{Conclusion}\label{conclusion}
	
	This work quantified avoided transmission due to local renewable power generation using the newly developed method of MPI-MPE curves that plot annual maximum power imports and exports as a function of installed renewable generation capacity. Two case studies were analyzed, with varying PV size and battery capacity, and at two sample locations in Germany and Japan. It was found that avoided transmission can occur, depending on the specific set-up. Without battery, transmission can only be avoided at the considered location in Japan, assuming moderate PV penetration. In contrast, no transmission avoidance was observed in the analyzed German location. With the installation of batteries, however, avoided transmission was observed in both locations. This, however, comes at quite high investment cost which is most probably not justified by achievable cost savings from avoided transmission.
	
	The MPI-MPE diagram developed in this work proved to be an effective and promising method to estimate avoided transmission effects. The approach can easily be applied to wider variety of load and generation profiles for different community set-ups of interest. The approach presented here relies on some simplifications which can lead to overestimating avoided transmission.  Perfect foresight of both PV generation and load is assumed, ommitting the effects of uncertainty. Besides, the work presented here employed the load profile data of one specific year combined with average weather data, applied to a community with PV generation. Since the distribution pattern of load and solar irradiation peaks is different in each year, the degree of avoided transmission effect might be more precisely determined by using weather input data of several years. Also, this study investigated the effect in only two	communities. Applying the method to other communities with different configurations and input data would lead to more precise and general knowledge about the conditions of avoided transmission effects. Finally, adding the energy perspective of electricity exchange between the community and the public grid could provide insights that help evaluating the value of battery storage better. Notwithstanding, the results of such more detailed studies could be very useful for policy makers and DSOs at the stage of infrastructure planning.
	
	%	\section*{Acknowledgment}

	%	The authors would like to thank...

	%	\ifCLASSOPTIONcaptionsoff
	%	\newpage
	%	\fi

	
	
	
	\bibliographystyle{IEEEtran}
	\bibliography{IEEEabrv,library}

	
	%	\begin{IEEEbiography}{Shigeyoshi Sato}
	%		Biography text here.
	%	\end{IEEEbiography}
	
	%	\begin{IEEEbiographynophoto}{Anke Weidlich}
	%		Biography text here.
	%	\end{IEEEbiographynophoto}
	
	%\newpage

	
	%\vfill
	

\end{document}